1) Title: Stress-induced hyperalgesia instead of analgesia in patients with chronic musculoskeletal pain

2) Running head: Stress-induced pain modulation in chronic pain

3) Authors: Martin Löffler[1,2,3], Pia Schneider[1], Sigrid Schuh-Hofer[4], Sandra Kamping[1,6], Katrin Usai[1], Rolf-Detlef Treede[4], Frauke Nees[1,5]*, Herta Flor[1]*
*both authors contributed equally to this work.

4) Affiliations:
[1]Institute of Cognitive and Clinical Neuroscience, Central Institute of Mental Health, Medical Faculty Mannheim, Heidelberg University, Germany
[2]Integrative Spinal Research Group, Department of Chiropractic Medicine, Balgrist University Hospital, University of Zurich, Zurich, Switzerland
[3]University of Zurich, Zurich, Switzerland
[4]Chair of Neurophysiology, Mannheim Center for Translational Neurosciences, Medical Faculty Mannheim, Heidelberg University, Germany
[5]Institute of Medical Psychology and Medical Sociology, University Medical Center Schleswig Holstein, Kiel University, Kiel, Germany
[6]Tabea Hospital, Outpatient Clinic for Chronic Pain, Hamburg, Germany

5) Corresponding author:
Herta Flor, PhD, Institute of Cognitive and Clinical Neuroscience, Central Institute of Mental Health, Medical Faculty Mannheim, Heidelberg University, Germany, Square J5, 68159 Mannheim, Germany
Email: Herta.Flor@zi-mannheim.de
Telephone: +49-621-17036301, Fax: +49-621-17036305

6) Funding: This work was supported by grants of the Deutsche Forschungsgemeinschaft to FN and HF (SFB1158/B03), to FN (NE 1383/14-1), RDT (SFB1158/S01 and B09) and the Bundesministerium für Bildung und Forschung (01EC1940B) to HF and ML.

7) Declarations of interest: none

**Abstract**

Many individuals with chronic musculoskeletal pain (CMP) show impairments in their pain-modulatory capacity. Although stress plays an important role in chronic pain, it is not known if stress-induced analgesia (SIA) is affected in patients with CMP. We investigated SIA in 22 patients with CMP and 18 pain-free participants. Pain thresholds, pain tolerance and suprathreshold pain ratings were examined before and after a cognitive stressor that typically induces pain reduction (SIA). Whereas the controls displayed a significant increase in pain threshold in response to the stressor, the patients with CMP showed no analgesia. In addition, increased pain intensity ratings after the stressor indicated hyperalgesia (SIH) in the patients with CMP compared to controls. An exploratory analysis showed no significant association of SIA or SIH with spatial pain extent. We did not observe significant changes in pain tolerance or pain unpleasantness ratings after the stressor in patients with CMP or controls. Our data suggest that altered stress-induced pain modulation is an important mechanism involved in CMP. Future studies need to clarify the psychobiological mechanisms of these stress-induced alterations in pain processing and determine the role of contributing factors such as early childhood trauma, catastrophizing, comorbidity with mental disorders and genetic predisposition.

## Introduction

Stress has inhibitory (stress-induced analgesia, SIA) as well as facilitatory (stress-induced hyperalgesia, SIH) effects on pain (Reinhardt et al., 2013). Both, SIA and SIH, are part of an in-built defense system, which serves to protect the body from injury (Butler and Finn 2009; Fanselow 1986). To what extent stress induces hyper- or hypoalgesic effects, depends on the type of stressor, the type of pain assessment and interindividual variability.

Although acute stress is usually associated with analgesic effects (Costa et al., 2005), it may induce hyperalgesia in healthy humans under specific conditions. For example, cognitive stressors have been linked to hypoalgesia (Flor et al., 2002; Flor and Grüsser 1999; Yilmaz et al., 2010), whereas stressors including social feedback often induce hyperalgesia (Cathcart et al., 2008; Cathcart et al., 2010; Cathcart et al., 2009; Crettaz et al., 2013; Ferreira et al., 2018; López-López et al., 2021), with some exceptions (Ferreira et al., 2021).

SIA or SIH further depend on the outcome parameters of the experimental paradigm. In controls, the human studies on SIA show increased pain thresholds, pain tolerance (Flor et al., 2002; Flor and Grüsser 1999) and nociceptive withdrawal reflex thresholds (Willer et al., 1981) to painful electric stimulation and increased pain thresholds, pain tolerance and decreased suprathreshold pain ratings to painful mechanical stimulation (Yilmaz et al., 2010). Stressors typically associated with SIH reduce heat pain thresholds but not mechanical pain thresholds (Crettaz et al., 2013; Ferreira et al., 2018). Mixed results were found for wind-up ratio and cold pain thresholds (no change (Crettaz et al., 2013) versus increase (Ferreira et al., 2018)). Pressure pain thresholds did not change after cognitive plus social stress (Ferreira et al., 2018), and yielded mixed results (no change (Crettaz et al., 2013) versus decrease (López-López et al., 2021)) although the same stressor was used in both tasks. In summary, the current findings indicate that SIA occurs in humans primarily for mechanical and electric pain sensitivity, whereas SIH appears after heat, cold or pressure pain. In addition, both are subject to interindividual variability (Reinhardt et al., 2013).

Studies on patients with chronic pain are equivocal. Patients with chronic tension type headache showed stronger stress-induced decreases of pressure pain thresholds at the head (Cathcart et al., 2008; Cathcart et al., 2010) and cold pain thresholds and stronger increases of suprathreshold cold pain ratings (Cathcart et al., 2009), i.e. signs of increased SIH. However, patients with temporomandibular disorder showed decreases of wind-up ratio, i.e. signs of SIA, after a stressor that increased windup ratio in controls (Ferreira et al., 2018). Further, patients with fibromyalgia showed absent or delayed decreases of pressure pain thresholds and –tolerance after social stress that elicited SIH in controls (López-López et al., 2021), while another study showed decreases of pressure and cold pain thresholds after social stress in patients with fibromyalgia, but decreased heat pain thresholds in controls (Crettaz et al., 2013). All prior studies in patients with chronic pain have employed stress paradigms typically associated with SIH. Comparable studies investigating the effects of stressors that typically elicit SIA are lacking.

In this study, we aimed to investigate if SIA is reduced in patients with chronic musculoskeletal pain (CMP) using a purely cognitive stressor. We employed electric pain thresholds as primary outcome. As recommended (Lacourt et al., 2012), we additionally assessed pain tolerance and suprathreshold ratings as secondary outcomes.

We expected to find hypoalgesic effects in controls, but reduced hypoalgesia in patients with CMP. In an exploratory analysis we tested if SIA is reduced in patients with more widespread pain.

## Methods

*Participants*

Twenty-two patients (15 women, age: mean (M)=55.90 standard deviation (SD)=12.36, see Table 1 for demographic and clinical data) with chronic localized or widespread musculoskeletal back pain and 18 pain-free healthy controls (15 women, age: M=52.94, SD=11.93) participated in the study. To our knowledge, no prior study compared the SIA magnitude between patients with chronic pain and controls. We therefore estimated the required sample size for our study based on an inhibitory pain mechanism. A meta-analysis (Lewis et al., 2012) found an overall large effect size (hedges g: 0.78) for the difference between patients with chronic pain and controls in the magnitude of conditioned pain modulation. The confidence interval (95%: 0.48 to 1.08) ranged from medium (g > 0.5) to large effects (g > 0.8) and the overall effect size in studies measuring pain thresholds was medium (hedges g: 0.61; 95% confidence interval: 0.13 to 1.10). A study that employed electrical pain thresholds to compare conditioned pain modulation between patients with CMP and controls found a medium effect size (hedges g: 0.6) as well (Lautenbacher and Rollman 1997). We therefore carried out a power calculation (G*power (Faul et al., 2007)) based on a medium effect size. We estimated the minimal required sample size for an interaction effect (time*group) and the following values: alpha error=0.05; power (1-beta)=0.80; effect size f=0.25 (medium effect size), correlation among repeated measures r=0.50. The calculated total sample size required was found to be 34 (17 per group). Therefore, a minimum of 17 patients and 17 controls was required to detect a medium effect size for a time*group interaction effect.

All inclusion and exclusion criteria were verified by experienced pain clinicians. Inclusion criteria for the patients with CMP were (a) suffering from back pain (local or widespread) for more than 3 months and (b) a current episode of back pain lasting more than 4 weeks prior to the experiment. A representative assessment of patients with chronic back pain showed that the majority of these patients reported additional pain sites (Gerhardt et al., 2014). We therefore included patients with other (primary) pain sites, such as the neck, hips, upper legs or shoulders. Exclusion criteria for both groups were (a) infections, (b) disorders involving neurological symptoms, such as multiple sclerosis, diabetes or Parkinson's disease, (c) neurological symptoms such as loss of skin sensation, abnormal skin sensation or muscle weakness, (d) possible pregnancy, (e) cardiovascular problems or (f) brain injuries. Controls were additionally excluded if they (a) reported a pain episode (> 1 day per week spent in pain) during the 3 months prior to the experiment or a previous diagnosis of chronic pain, (b) used analgesic, anxiolytic or antidepressant medication in the twelve months prior to the experiment or (c) reported any comorbid mental disorders. Patients with CMP with comorbid anxiety and depression (but no other comorbid mental disorder) were included as these are characteristic for chronic pain. Similar to SIA, nonsteroidal anti-inflammatory drugs, as well as other analgesic drugs have been shown to target the descending pain control system (Vanegas et al., 2010). In agreement with their physician, we therefore asked patients to refrain from using nonsteroidal anti-inflammatory drugs for one day (Grosser et al., 2017) and to refrain from using muscle relaxants and acute analgesic medication other than nonsteroidal anti-inflammatory drugs for three days prior to the experiment (Mandelli et al., 1978).

The groups were matched for age and recruited through announcements in local newspapers, the institute's website, the outpatient pain clinic of the Institute of Cognitive and Clinical Neuroscience, general practitioners and physiotherapy practices. The Ethics Committee of the Medical Faculty Mannheim, University of Heidelberg, Germany, approved the study (reference number: 2010-263N-MA) and written informed consent was obtained from each participant. No participant terminated the experiment prematurely.

| | **Scale Range (min-max)** | **Patients with chronic musculoskeletal pain (CMP)** | **Healthy Controls (HC)** | **Group test Chi²/t(df), p, d** |
|---|---|---|---|---|
| **N (male/female)** | | 22 (7/15) | 18 (3/15) | Chi²(1)=0.54, p=0.54, d=0.23 |
| **Age: M±SD** | | 56.55±12.43 | 52.94±11.93 | t(36.45)=-0.76, p=0.54, d=0.30 |
| **HADS Anxiety: M±SD** | 0-21 | 7.09±3.12 | 5.28±2.22 | t(34.67)=-2.05, p=0.08, d=0.67 |
| **HADS Depression: M±SD** | 0-21 | 13.32±2.44 | 13.48±1.64 | t(35.28)=0.35, p=0.73, d=0.08 |
| **PRSS Catastrophizing: M±SD** | 0-5 | 1.54±0.95 | 0.41±0.42 | **t(30.81)=-3.96, p=0.003, d=1.53** |
| **PRSS Active Coping: M±SD** | 0-5 | 3.08±0.79 | 4.14±0.73 | **t(36.33)=3.30, p=0.008, d=1.39** |
| **TICS Stress: M±SD** | 0-4 | 1.54±0.58 | 0.84±0.38 | **t(35.58)=-2.60, p=0.03, d=1.37** |
| **Pain intensity 4 weeks prior to experiment: M±SD** | 0-10 | 4.90±1.65 | n.a. | n.a. |
| **MPI pain severity: M±SD** | 0-6 | 3.05±1.30 | n.a. | n.a. |
| **MPI interference: M±SD** | 0-6 | 2.34±1.69 | n.a. | n.a. |
| **MPI affective distress: M±SD** | 0-6 | 2.38±0.73 | n.a. | n.a. |
| **MPI support: M±SD** | 0-6 | 2.63±1.28 | n.a. | n.a. |
| **MPI self-control: M±SD** | 0-6 | 2.26±1.64 | n.a. | n.a. |

| **WPI Widespread Pain: M±SD** | 0-19 | 6.50±4.10 | n.a. | n.a. |
|---|---|---|---|---|
| **SS Symptom Severity: M±SD** | 0-12 | 4.07±3.36 | n.a. | n.a. |
| **CPG Chronic Pain Grade: Median (IQR)** | 0-4 | 2(1-3) | n.a. | n.a. |

*Table 1: Demographic and clinical characteristics of the patients and controls. Significant differences (p<0.05) are depicted in bold. HADS: Hospital Anxiety and Depression Scale, PRSS: Pain-Related Self Statements Scale, TICS: Trier Inventory of Chronic Stress; MPI: Multidimensional Pain Inventory; WPI: Widespread Pain Index, SS: Symptom Severity Scale, CPG: Chronic Pain Grade, M: mean, SD: standard deviation, df: degrees of freedom, IQR: Interquartile range; n.a.: not applicable.*

### *Pain-related variables*

Prior to the experiments, the participants completed the Chronic Pain Grade (Von Korff et al., 1992), the Fibromyalgia Symptom Scale (Wolfe et al., 2010), the Pain-Related Self Statements Scale (Flor et al., 1993) as a measure of catastrophizing and the West Haven-Yale Multidimensional Pain Inventory (Flor et al., 1990; Kerns et al., 1985). Mean pain intensity four weeks prior to the experiment was assessed using a numeric rating scale from 0=”no pain” to 10=”worst pain imaginable” and the onset of the chronic pain condition was assessed using a categorical scale (less than 3 months, 3 to 12 months, 1 to 5 years, 5 to 10 years, more than 10 years) (Nagel et al., 2002). The Chronic Pain Grade (Von Korff et al., 1992) is a seven-item instrument that measures chronic pain severity in the two dimensions intensity and disability. It classifies patients into five hierarchical grades: Grade 0 (pain free), Grade I (low disability–low intensity), Grade II (low disability–high intensity), Grade III (high disability–moderately limiting) and Grade IV (high disability–severely limiting). It was found to be valid (convergent validity ranging from r=0.28 to r=0.84, depending on the comparison scales; factor loadings ranging from 0.77 to 0.89) and reliable (α=0.91) for use in a general population as a self-completion questionnaire (Smith et al., 1997). The Fibromyalgia Symptom Scale (Wolfe et al., 2010) determines the severity of fibromyalgia symptoms on two different scales, the Widespread Pain Index and the Symptom Severity Scale. The Widespread Pain Index assesses pain or tenderness in 19 body regions. When summed, these items result in a score between 0 and 19. The Symptom Severity Scale assesses additional somatic, cognitive and affective symptoms related to fibromyalgia on categorical items. When summed, these items result in a score between 0 and 12. For the diagnosis of fibromyalgia a combination of the two scales of the Fibromyalgia Symptom Scale was shown to have a sensitivity of 96.6% and a specificity of 91.8%, where fibromyalgia was defined for patients with scores of either Widespread Pain Index > 7 and Symptom Severity Scale > 5 or Widespread Pain Index >= 3 and Symptom Severity Scale > 9 (Wolfe et al., 2011). The Pain-Related Self Statements Scale (Flor et al., 1993) assesses catastrophizing and active coping and is a German language equivalent of the Coping Strategies Questionnaire (Rosenstiel and Keefe 1983). It has excellent reliability (α=0.92 for catastrophizing and α=0.88 for active coping) and validity as shown by significantly higher values for pain catastrophizing and significantly lower values for active coping in pain patients compared to healthy controls, and low to moderate correlations with other pain-related variables such as amount of daily activities, affective distress or pain severity. The West Haven-Yale Multidimensional Pain Inventory assesses patients' pain severity, interference of the pain, self-control, negative mood and social support, spouse responses and daily activities. The West Haven-Yale Multidimensional Pain Inventory has been used in a large number of studies with diverse samples of chronic pain patients and has excellent reliability and validity (Flor et al., 1990; Kerns et al., 1985).

*Assessment of stress, mood and comorbidity*

Prior to the experiments, the participants completed the Hospital Anxiety and Depression Scale (Snaith and Zigmond 1994), and a short version of the Trier Inventory of Chronic Stress (Schulz et al., 2004). The Trier Inventory of Chronic Stress (Schulz et al., 2004) measures chronic stress with a mean score of six scales (work overload, worries, social stress, lack of social recognition, work discontent, and intrusive memories). The answers are recorded on a 0–4 rating scale, with a total number of 30 items. The scale is validated in German participants and has a reliability between α=0.84 and α=0.91. The Hospital Anxiety and Depression Scale (Snaith and Zigmond 1994) assesses anxiety and depressive symptoms on 7 items each. The scale has excellent reliability (α ranges from 0.78–0.93 for the anxiety subscale and from 0.82–0.90 for the depression subscale). It was found to perform well in assessing the symptom severity and caseness of anxiety disorders and depression in both, somatic and mental disorders, and primary care patients as well as in the general population (Bjelland et al., 2002).

To further assess comorbid mental disorder, all participants were interviewed by a trained psychologist using the German version of the Structured Clinical Interviews for DSM IV (SCID I and II) (Fydrich et al., 1997; Wittchen et al., 1997).

*Experimental procedure*

All participants were examined on two separate days. On the screening day, the difficulty of the cognitive stressor (see below) was determined to ensure successful stress induction during the SIA experiment and the SCID interview was carried out. Between the screening day and the experimental day patients completed all questionnaires at home. The screening day and the experimental days were separated by one week. On the experimental day, the SIA experiment was carried out. Participants were seated in a chair and electric pain thresholds, pain tolerance and suprathreshold pain ratings were assessed before and after stress induction. The order of the tests in the stress paradigm was as follows: 1) threshold determination, 2) test trials to calibrate stimulus intensity for suprathreshold pain testing, 3) suprathreshold pain trials 4) baseline physiological recordings, 6) Stress induction 7) suprathreshold pain trials, 8) threshold determination. For a graphical depiction of the order of the experimental procedure see Figure 1 and see below for details on each test.

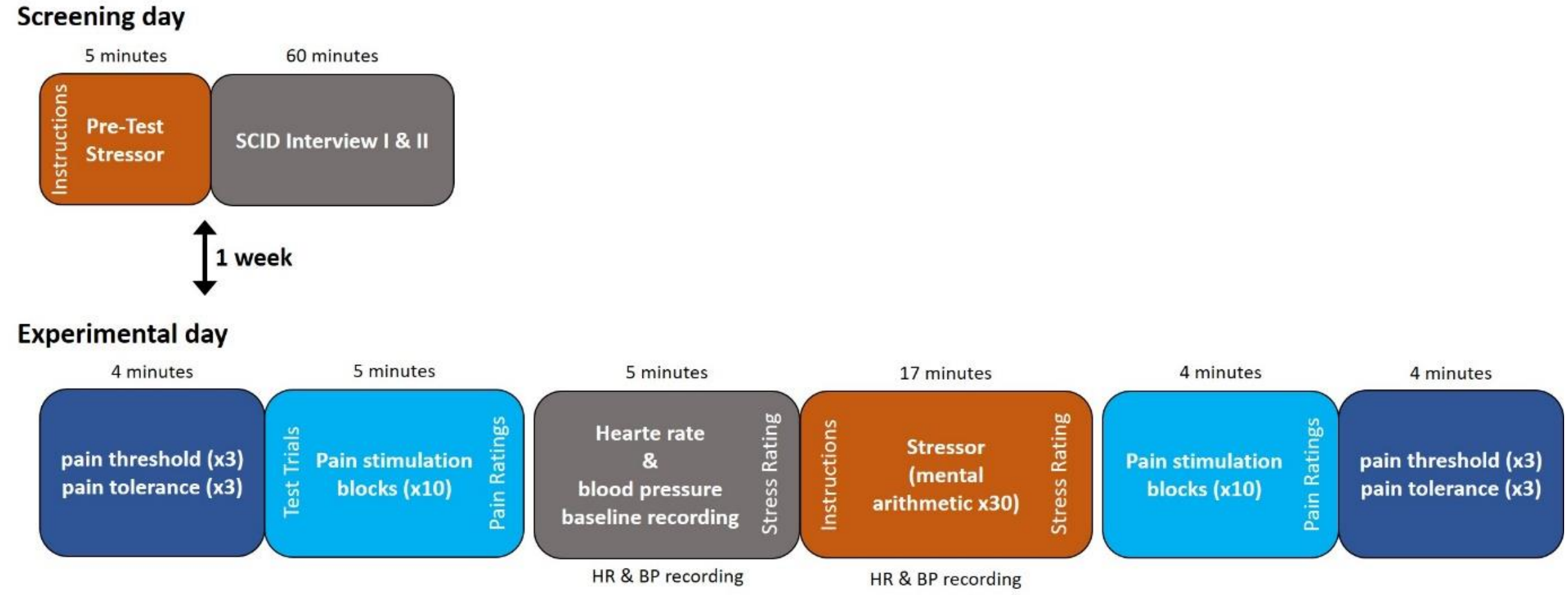

*Figure 1: Experimental procedure: On the screening day, all participants underwent a pretest to determine the individual difficulty of the stress task, followed by the SCID interview. The experimental day was scheduled one week after the screening day. At the beginning of the experimental day, pain threshold and pain tolerance were measured in three ascending series followed by a series of 10 stimuli with an intensity corresponding to 50% of the pain tolerance intensity. These 10 stimuli were rated on a numeric rating scale with respect to pain intensity (0 = not painful; 100 = worst pain imaginable) and unpleasantness (0 = not unpleasant; 100 = extremely unpleasant). Before and during stress induction, blood pressure and heart rate data were gathered. Before and after the stress induction, subjects were asked to rate the perceived stress before or during stress induction respectively on a numeric rating scale (0 = no stress; 100 = extreme stress). Finally, as post-stress measurement, application and rating (intensity and unpleasantness) of 10 painful stimulus blocks and pain threshold and pain tolerance determination were repeated. HR: heart rate; BP: blood pressure; SCID: Structured Clinical Interview I and II for DSM IV.*

### *Pain threshold, pain tolerance and suprathreshold pain testing*

Pain perception thresholds, pain tolerance and suprathreshold pain ratings were measured before and after the stressor. For electric stimulation, a pair of needle electrodes (20 millimeter long, 0.35 millimeter uninsulated tip, 2 millimeter² stimulation area, model: 9013R0272, 28G, Alpine Biomed ApS, Skoviunde, Denmark) was placed subcutaneously at the left lower back, 2 centimeter lateral to the spine, between L1 and L3 (1 millimeter needle separation) while subjects were leaning forward, and the subjects then settled back comfortably into the chair. This site was used for all participants to stimulate at a body region close to the current clinical pain, while keeping the stimulation site constant across participants (21 out of 22 patients had pain at the lower back). Needle electrodes were placed fifteen minutes before starting the experimental procedure to minimize potential effects of pain during needle placement. Electrodes were fixated using surgical tape. Electric stimuli were applied using a constant current stimulator (model DS7A; Digitimer, Hertfordshire, England). Needle electrodes were chosen to limit stimulation to a specific fibre type, namely Aδ fibers and to elicit a rapid and sharp painful sensation (Inui et al., 2002). The experiment was performed using Presentation® software (Version 14.0, http://www.neurobs.com/).

During the threshold determination participants received trains of electric stimuli (8 stimuli of 2ms stimulus duration with 112msec inter stimulus interval) every 2 seconds. Stimulus intensity started at 0mA for each trial and was increased manually between the trains by the experimenter (PS). Participants were instructed to press a button when they felt that the stimulus had become just noticeably painful (pain threshold) and when they could no longer tolerate a higher stimulus intensity (pain tolerance). Each measure was determined three times before the stressor and three times after the stressor.

After the baseline threshold determination, the stimulation intensity for suprathreshold pain testing was determined. The first threshold and tolerance assessment was discarded and 50 percent of the difference between pain threshold and pain tolerance were added to the pain threshold, to calculate a preliminary stimulation intensity for a first test trial. The stimulation intensity was then calibrated to a perceived pain intensity rating of 50 on a numeric rating scale (endpoints 0="no pain" and 100="worst pain imaginable"). For that purpose the perceived pain intensity was assessed after each test trial. Test trials (duration 11.8 seconds) consisted of 105 stimuli of 2 milliseconds duration with an inter stimulus interval of 112 milliseconds. The stimulation intensity was adapted between the test trials to reach a pain intensity rating of 50 out of 100 points, or to reach a rating closest possible to 50. The resulting stimulation intensity of each subject was used for suprathreshold pain tests.

During the suprathreshold pain trials, the participants received painful stimulation equal to the last test trial. Before and after the stressor 10 stimulation trials were administered with off blocks of 11.8 seconds duration between each trial. The mean levels of perceived pain intensity and pain unpleasantness were assessed after the 10 stimulation blocks using numeric rating scales with the endpoints 0 ("no pain" / "not unpleasant") and 100 ("worst pain imaginable" / "extremely unpleasant").

In total, four indices of stress-induced pain modulation were derived: (1) mean pain threshold, (2) mean pain tolerance, (3) pain intensity ratings and (4) pain unpleasantness ratings by subtracting the respective values before the stressor from the values after the stressor. An increase of pain threshold or tolerance, as well as a decrease of suprathreshold pain ratings after the stressor would indicate SIA, while a decrease of pain threshold and tolerance as well as an increase of suprathreshold pain ratings would indicate SIH.

*Stress induction*

The stressor used in this study was mental arithmetic. In order to increase the stressfulness of the task, white noise was presented continuously and increased from 65 to 80 decibel from the first to the last arithmetic calculation. This particular stressor was chosen because it previously induced SIA in healthy participants (Flor et al., 2002; Flor and Grüsser 1999; Yilmaz et al., 2010). The mental arithmetic tasks were similar to those from the Konzentrations- und Leistungstest (Concentration and Performance Test (Düker and Lienert 1959; Lukesch and Mayrhofer 2001)) and were presented by a female voice via earphones. For each task participants first consecutively solved two sets of arithmetic tasks with three numbers (e.g. task one: 5+9+4=**18** and task two: 3+8+11=**22**). In a second step, participants combined the results of the two tasks using two rules: (1) If the result of the second task was smaller than the result of the first task, the result of the second task had to be subtracted from the result of the first task, (2) if the result of the first task was smaller than that of the second task, the result of the second task had to be added to the result of the first task (i.e. task results are added in the example: 18+22=**40**). As a last step, an additional third arithmetic operation was announced via earphones and had to be executed on the interim result of the second step (e.g. **40***2=80). Afterwards, participants verbally reported the final result (i.e. 80 in the example). Each task had to be solved within 30 seconds. In total subjects had to solve 30 tasks, resulting in a duration of the stressor of 15 minutes.

To account for individual cognitive performance, five parallel versions of the mental arithmetic task with varying difficulty (based on the arithmetic operations) were prepared. The individual difficulty level was determined for each participant on a separate day. For that purpose, 5 tasks of the lowest level of difficulty were presented. If the participant solved at least three of those, the next level was presented. If the participant solved at least three tasks at the second level again, the next level was presented. This was continued until the participant made more than one error within a level or the highest level was reached. In the main experiment one level above the resulting difficulty level was used for stress induction. The perceived stressfulness of the task was assessed before and after the stressor using a numeric rating scale with the endpoints 0 (“no stress”) and 100 (“extreme stress”).

*Physiological stress response: heart rate and blood pressure*

To assess physiological effects of the stressor, blood pressure and heart rate were measured with the Criticare 506N vital signs monitor (Criticare Systems Inc.,

Waukesha, USA), using a sampling rate of one per minute. Heart rate and blood pressure were measured during a 5 minute resting interval immediately before the stress phase and throughout the 15 minutes of the stress phase. To estimate the physiological stress response we averaged heart rate, diastolic and systolic blood pressure across the measurements before the stressor and during the stressor for all further analyses. We chose blood pressure and heart rate instead of cortisol levels as physiological stress outcomes, because SIA with a cognitive stressor has previously been found to be independent of hypo-thalamic-pituitary-adrenal axis functioning (Lewis et al., 1981), which is more closely related to stress in a social situation (Choi et al., 2012).

## Statistical analysis

Statistical analyses were performed using RStudio 1.0.143 (RStudio, Inc.) with R 3.4.0 (R_Core_Team 2017). Outliers, which were defined as values more than two standard deviations above or below the respective mean, were removed (maximum percentage of outliers per variable: controls=11%, patients=9%). Missing values were imputed using the MICE package in R, applying predictive mean matching for numeric variables and a proportional odds model for ordered variables (maximum percentage of imputed values per variable: controls=17%, patients=14%). Missing values were mainly found for heart rate and blood pressure (see Supplementary Figure 1), due to technical problems in two participants. For all statistical analyses, $\alpha$ was set to 0.05. We used the false discovery rate (Benjamini and Hochberg 1995) to adjust for multiple comparisons in all analyses. In case of violations of the sphericity assumption, Greenhouse–Geisser correction was applied. Effect sizes are reported as Cohen's d (Cohen 1992) for post-hoc t-tests and as generalized $\omega^2$ for analyses of variances, as previously recommended (Lakens 2013), especially for sample sizes below n=50 (Albers and Lakens 2018).

*Sample characteristics and comparisons*

Gender, age, the Hospital Anxiety and Depression Scale anxiety and depression subscales, the Pain-Related Self Statements Scale Catastrophizing and Active Coping subscales and the Trier Inventory of Chronic Stress scale were compared between healthy controls and patients with CMP using a chi-square test or t-tests (corrected using the false discovery rate (Benjamini and Hochberg 1995)).

*Manipulation checks: Psychophysiological stress response: perceived stress, heart rate and blood pressure*

To check successful stress induction and for differences in stress responsivity in patients and controls we compared the difficulty of the tasks used for stress induction using a Wilcoxon rank-sum test and the number of correctly solved arithmetic tasks using a t-test. Further, we computed separate analyses of variance for the perceived stress ratings, systolic and diastolic blood pressure and heart rate with group (healthy versus patient) as between and time (baseline versus post) as within effects. We used pairwise post-hoc t-tests (corrected using the false discovery rate (Benjamini and Hochberg 1995)) to compare stress measures within and between groups.

*The effect of stress on primary and secondary outcomes*

For pain thresholds, suprathreshold pain intensity and pain unpleasantness ratings and pain tolerance, separate analyses of variance were computed with group (healthy versus patient) as between and time (baseline versus post) as within factors. We used pairwise post-hoc t-tests (corrected using the false discovery rate (Benjamini and Hochberg 1995)) to compare the SIA measures within and between groups. To test if any effects of time or time by group interactions were related to age or gender we additionally computed the same analysis of variance including the main effect and interaction effects of gender and age as between subject factor or covariate, respectively (see Supplementary Table 1). Since we found a significant time by group interaction for systolic and diastolic blood pressure, with the strongest effect for systolic blood pressure (see below and Table 3), we additionally added systolic blood pressure as a covariate to the model (see Supplementary Table 2).

*Exploratory analyses: spatial pain extent*

A possible relationship of altered SIA in patients with CMP with the extent of their clinical pain was examined using correlations of the Widespread Pain Index with the difference in pain thresholds, pain tolerance and ratings during the experiment.

## Results

*Characterization and comparison of the samples*

Patients with CMP showed significantly higher values for pain catastrophizing (t(30.81)=-3.96, p=0.003, d=1.53) and life-time stress exposure (t(35.58)=-2.60, p=0.03, d=1.37) and lower scores on the Pain-Related Self Statements Scale active coping scale (t(36.33)=3.30, p=0.008, d=1.39). Age (t(36.45)=-0.76, p=0.54, d=0.30), gender (Chi$^2$(1)=0.54, p=0.54, d=0.23), anxiety (t(34.67)=-2.05, p=0.08, d=0.67) and depression (t(35.28)=0.35, p=0.73, d=0.08) scores did not significantly differ between patients with CMP and controls, see Table 1.

Based on the diagnostic criteria of Wolfe et al. (2010) nine of the 22 patients fulfilled the diagnostic criteria for fibromyalgia. In the Widespread Pain Index, twenty-one of 22 patients reported lower back pain, 11 of those 21 additionally complained of upper back pain and one subject only suffered from upper back pain. Back pain was named as the worst pain by 14 out of 22 patients. Most patients had been suffering from chronic musculoskeletal back pain for more than 10 years (13 patients), 3 patients for more than 5 years, 4 patients for more than one year and 2 patients for more than 3 months.

One patient fulfilled the criteria of an acute depressive episode and was referred to our outpatient clinic after the experiment. Three patients reported a single depressive episode in the past. No healthy participant reported any axis I or II disorders.

Neither patients nor controls reported the current use of antidepressants or anxiolytic drugs. Eight patients regularly used nonsteroidal anti-inflammatory drugs. Past medication was not reported by any control person, but two patients (patient 1: transdermal opioid patch for 4 weeks, 13 years prior to the experiment; patient 2: oral use of an opioid for 6 weeks, 15 years prior to the experiment).

*Manipulation checks: Psychophysiological stress response: perceived stress, heart rate and blood pressure*

Task difficulty did not significantly differ ($W=237.5$, $p=0.27$, $d=0.34$) between CMP patients (median=1.5, interquartile range = 1-3) and controls (median=2.5, interquartile range = 1.25-3). Further, control subjects ($M = 9.92$, $SD = 5.10$) and CMP patients ($M = 8.29$, $SD = 5.19$) did not significantly differ in the number of correctly solved arithmetic tasks ($t(32.19)=0.93$, $p=0.36$, $d= 0.32$).

Perceived stress increased significantly after the stressor compared to before (time: $F(1,37)=71.26$, $p<0.001$, $\omega^2=0.430$) and did not significantly differ between healthy participants and patients (group: $F(1,37)=2.27$, $p=0.14$, $\omega^2=0.010$; time*group: $F(1,37)=3.02$, $p=0.09$, $\omega^2=0.012$). Heart rate, systolic and diastolic blood pressure significantly increased during the stressor compared to before the stressor (time: heart rate: $F(1,37)=11.26$, $p=0.002$, $\omega^2=0.071$; systolic blood pressure: $F(1,37)=30.68$, $p<0.001$, $\omega^2=0.103$; diastolic blood pressure: $F(1,37)=15.99$, $p<0.001$, $\omega^2=0.070$). Heart rate, systolic and diastolic blood pressure did not significantly differ between patients and controls (group: heart rate: $F(1,37)=0.37$, $p=0.54$, $\omega^2=-0.011$; systolic blood pressure: $F(1,37)=0.61$, $p=0.44$, $\omega^2=-0.007$; diastolic blood pressure: $F(1,37)=0.84$, $p=0.36$, $\omega^2=-0.003$). However, the increase in systolic (time*group: $F(1,37)=6.79$, $p=0.01$; $\omega^2=0.020$) and diastolic (time*group: $F(1,37)=5.32$, $p=0.03$, $\omega^2=0.020$) blood pressure was significantly higher in patients than controls. This was not the case for heart rate during the stressor (time*group: $F(1,37)=0.41$, $p=0.53$, $\omega^2=-0.004$), see Table 3.

| | Patients with chronic musculoskeletal pain (CMP) | | | | Healthy Controls (HC) | | | |
|---|---|---|---|---|---|---|---|---|
| | **Before (M±SD)** | **During (M±SD)** | **t(df); p** | | **Before (M±SD)** | **During (M±SD)** | **t(df); p** | |
| **Heart Rate** | 66.35±11.11 | 73.24±9.47 | **t(21)=-2.92; p=0.02; d=0.62** | | 65.77±10.59 | 70.41±9.59 | t(17)=-1.83; p=0.08; d=0.43 | |
| **Blood pressure: systolic** | 128.22±14.65 | 144.02±16.52 | **t(21)=-6.55; p<.001; d=1.65** | | 129.81±16.51 | 135.12±16.23 | t(17)=-1.53; p=0.14; d=0.36 | |
| **Blood pressure: diastolic** | 74.19±9.24 | 82.53±9.23 | **t(21)=-5.35; p<.001; d=1.39** | | 74.86±10.99 | 76.84±8.73 | t(17)=-0.82; p=0.42; d=0.19 | |
| **Perceived stress (NRS)** | 17.11±21.56 | 64.21±22.50 | **t(21)=-8.03; p<.001; d=1.84** | | 14.19±21.23 | 51.56±27.19 | **t(17)=-4.71; p<.001; d=1.18** | |
| | **Before (M±SD)** | **After (M±SD)** | **t(df); p** | | **Before (M±SD)** | **After (M±SD)** | **t(df); p** | |
| **Pain Threshold (in mA)** | 1.23±0.68 | **1.30±0.72 †** | t(21)=-0.51; p=0.62; d=0.11 | Ø | 1.75±1.21 | 2.62±1.82 | **t(17)=-2.56, p=0.04; d=0.60** | ↓ |
| **Pain Tolerance (in mA)** | **2.23±1.33 †** | **2.15±1.09 †** | t(21)=0.31; p=0.76; d=0.07 | Ø | 3.99±3.23 | 3.76±2.61 | t(17)=0.31; p=0.76; d=0.07 | Ø |
| **Perceived pain intensity (NRS)** | 39.55±17.65 | **54.55±16.03 †** | **t(21)=-3.07; p=0.01; d=0.66** | ↑ | 41.28±14.90 | 38.72±18.65 | t(17)=0.75; p=0.47; d=0.18 | Ø |
| **Perceived pain unpleasantness (NRS)** | 45.00±23.90 | 50.45±24.10 | t(21)=-1.49; p=0.30; d=0.32 | Ø | 49.33±20.94 | 46.11±17.90 | t(17)=0.74; p=0.47; d=0.17 | Ø |

*Table 2: Change in perceived stress levels and associated changes in heart rate and blood pressure. The table depicts the stress ratings, systolic and diastolic blood pressure and heart rate before and during the stressor (upper part of the table) and before and after the stressor (lower part of the table). All values are depicted as mean (M) and standard deviation (SD). There were no significant differences between the patients and controls before or after the stressor for perceived stress ratings, blood pressure or heart rate (all t<2.00; p>0.11). Significant differences between the patients and controls before or after the stressor for thresholds or suprathreshold pain measures are highlighted in bold and using a cross (†p<0.05). Arrows indicate analgesic (green arrows pointing downwards), hyperalgesic (red arrows pointing upwards), or no significant effects (Ø) of the stressor. Within-subject significance levels are highlighted in bold; NRS: Numerical Rating Scale, df: degrees of freedom.*

*Differences between pain patients and healthy individuals in stress-induced analgesia*

We found a significant time*group interaction effect for pain thresholds (time*group: $F(1,37)=5.63$, $p=0.02$, $\omega^2=0.021$), indicating that the increase of pain thresholds in controls was reduced for patients with CMP. Post-hoc comparisons of pain thresholds before and after the stressor revealed a significant SIA effect for the healthy participants ($t(17)=-2.56$, $p=0.04$, $d=0.60$), but not for the patients with chronic pain ($t(21)=-0.51$, $p=0.62$, $d=0.11$), see Figure 2a and Table 3.

Further, a significant time*group effect for pain intensity ratings (time*group: $F(1,37)=7.76$, $p=0.008$, $\omega^2=0.052$), showed that patients with CMP displayed a stronger increase of pain intensity ratings than controls. Post-hoc comparisons of pain intensity ratings before and after the stressor revealed a significant SIH effect for patients with chronic pain ($t(21)=-3.07$, $p=0.01$, $d=0.66$), but not for healthy participants ($t(17)=0.75$, $p=0.47$, $d=0.18$), see Figure 2c and Table 3.

Pain tolerance showed a significant main effect for group (group: $F(1,37)=8.08$, $p=0.007$, $\omega^2=0.118$), with higher pain tolerance for the healthy participants than the patients with CMP. We found no significant effect for time (time: $F(1,37)=0.16$, $p=0.69$, $\omega^2=-0.005$) or the interaction of time*group (time*group: $F(1,37)=0.05$, $p=0.83$, $\omega^2=-0.006$) for pain tolerance, see Figure 2b and Table 3.

For pain unpleasantness ratings, we observed no significant effects for time (time: $F(1,37)=0.30$, $p=0.58$, $\omega^2=-0.003$), group (group: $F(1,37)=0.00$, $p=0.99$, $\omega^2=-0.022$) or time*group (time*group: $F(1,37)=2.36$, $p=0.13$, $\omega^2=0.006$), see Figure 2d and Table 3.

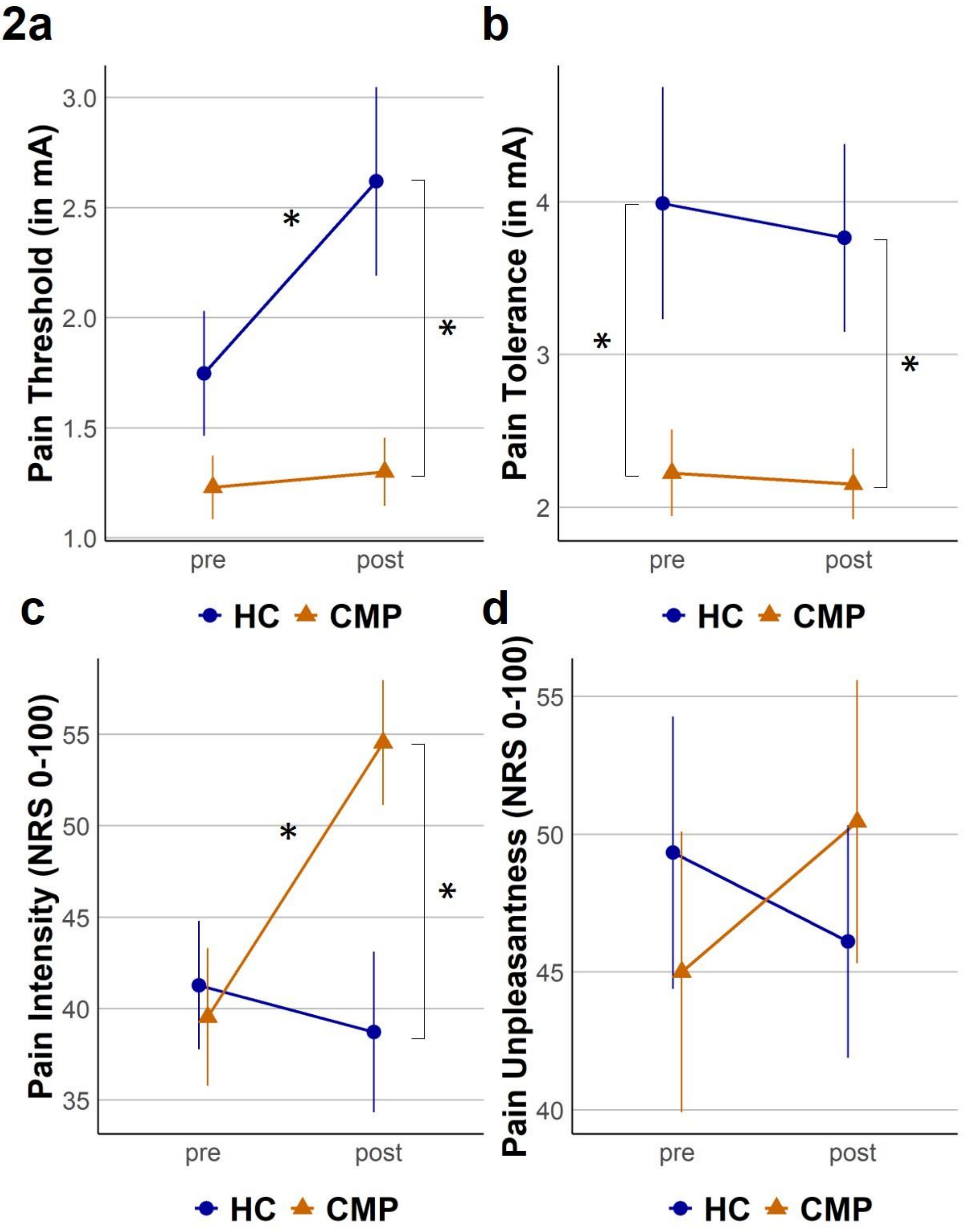

*Figure 2: Group differences in SIA. Thresholds and ratings before and after the stressor are depicted as mean and standard error of the mean. Patients with chronic musculoskeletal pain (CMP) are depicted with triangles and orange lines, healthy controls (HC) are depicted with circles and blue lines. Significant post hoc (corrected using the false discovery rate) differences are labeled with asterisks (*p<0.05). Figure 2a (top left) shows electric pain threshold before (pre) and after (post) the stressor, Figure 2b (top right) shows electric pain tolerance before (pre) and after (post) the stressor, Figure 2c (bottom left) shows pain intensity ratings in response to the test stimulus before (pre) and after (post) the stressor, Figure 2d (bottom right) shows pain unpleasantness ratings in response to the test stimulus before (pre) and after (post) the stressor.*

When we added age or gender as individual covariates to the analysis of variance models on primary and secondary outcomes, we did not find any significant interaction effects of the factor gender (all $F<3.52$, $p>.07$, $\omega^2<0.020$). For the covariate age we found a significant interaction effect age*time*group ($F(1,37)=4.31$, $p=0.047$, $\omega^2=0.010$) for the unpleasantness ratings, indicating that the stress-induced reduction of pain unpleasantness ratings became less with age in healthy participants but not in patients with CMP (see Supplementary Figure 1). There were no other significant interaction effects of the covariate age with the change in primary or secondary outcome measures after the stressor (all $F<1.33$, $p>0.26$, $\omega^2<0.002$), see Supplementary Table 1.

After adding systolic blood pressure as a covariate to the model, we did not find any significant interaction effects of systolic blood pressure with the change in primary or secondary outcome measures after the stressor (all $F<3.34$, $p>0.08$, $\omega^2<0.018$), see Supplementary Table 2. This indicates that the time*group interaction effects of pain thresholds and pain intensity ratings were not moderated by autonomic stress responsivity.

*Association of stress-induced analgesia and spatial pain extent*

We could not find significant correlations of changes in primary or secondary outcomes after the stressor with the extent of clinical pain (pain thresholds: $r(22)=-0.03$, $p=0.89$; pain tolerance: $r(22)=0.01$, $p=0.97$; pain intensity ratings: $r(22)=-0.35$, $p=0.25$; or pain unpleasantness ratings: $r(22)=-0.21$, $p=0.53$).

| | Time | group | Time*group | HC: pre/during | CMP: pre/during | Pre: HC/CMP | During: HC/CMP |
|---|---|---|---|---|---|---|---|
| **Heart Rate** | **F(1,37)=11.26; p=0.002; $\omega^2$=0.071** | F(1,37)=0.37; p=0.54; $\omega^2$=-0.011 | F(1,37)=0.41, p=0.53; $\omega^2$=-0.004 | t(17)=-1.83; p=0.08; d=0.43 | **t(21)=-2.92; p=0.02; d=0.62** | t(37.06)=-0.17; p=0.87; d=0.05 | t(36.25)=-0.93; p=0.72; d=0.30 |
| **Blood pressure: systolic** | **F(1,37)=30.68; p<0.001; $\omega^2$=0.103** | F(1,37)=0.61; p=0.44; $\omega^2$=-0.007 | **F(1,37)=6.79; p=0.01; $\omega^2$=0.020** | t(17)=-1.53; p=0.14; d=0.36 | **t(21)=-6.55; p<.001; d=1.65** | t(33.58)=0.18; p=0.86; d=0.06 | t(36.47)=-1.60; p=0.24; d=0.51 |
| **Blood pressure: diastolic** | **F(1,37)=15.99; p<0.001; $\omega^2$=0.070** | F(1,37)=0.84; p=0.36; $\omega^2$=-0.003 | **F(1,37)=5.32; p=0.03; $\omega^2$=0.020** | t(17)=-0.82; p=0.42; d=0.19 | **t(21)=-5.35; p<.001; d=1.39** | t(32.78)=0.15; p=0.88; d=0.05 | t(36.82)=-1.84; p=0.15; d=0.58 |
| **Perceived stress (NRS)** | **F(1,37)=71.26; p<0.001; $\omega^2$=0.430** | F(1,37)=2.27; p=0.14; $\omega^2$=0.010 | F(1,37)=3.02; p=0.09; $\omega^2$=0.012 | **t(17)=-4.71; p<.001; d=1.18** | **t(21)=-8.03; p<.001; d=1.84** | t(32.16)=-0.40; p=0.69; d=0.14 | t(29.20)=-1.48; p=0.30; d=0.51 |

| | Time | group | Time*group | HC: pre/post | CMP: pre/post | Pre: HC/CMP | Post: HC/CMP |
|---|---|---|---|---|---|---|---|
| **Pain Threshold (in mA)** | **F(1,37)=6.56; p=0.01; $\omega^2$=0.025** | **F(1,37)=7.85; p=0.008; $\omega^2$=0.115** | **F(1,37)=5.63; p=0.02; $\omega^2$=0.021** | **t(17)=-2.56, p=0.04; d=0.60** | t(21)=-0.51; p=0.62; d=0.11 | t(25.62)=1.62; p=0.12; d=0.53 | **t(21.42)=2.90; p=0.02; d=0.96** |
| **Pain Tolerance (in mA)** | F(1,37)=0.16; p=0.69; $\omega^2$=-0.005 | **F(1,37)=8.08; p=0.007; $\omega^2$=0.118** | F(1,37)=0.05; p=0.83; $\omega^2$=-0.006 | t(17)=0.31; p=0.76; d=0.07 | t(21)=0.31; p=0.76; d=0.07 | **t(21.72)=2.18; p=0.04; d=0.72** | **t(21.82)=2.45; p=0.04; d=0.81** |
| **Perceived pain intensity (NRS)** | **F(1,37)=5.13; p=0.03; $\omega^2$=0.032** | F(1,37)=2.55; p=0.12; $\omega^2$=0.023 | **F(1,37)=7.76; p=0.008; $\omega^2$=0.052** | t(17)=0.75; p=0.47; d=0.18 | **t(21)=-3.07; p=0.01; d=0.66** | t(37.95)=0.34; p=0.74; d=0.11 | **t(33.77)=-2.84; p=0.02; d=0.91** |
| **Perceived pain unpleasantness (NRS)** | F(1,37)=0.30; p=0.58; $\omega^2$=-0.003 | F(1,37)=0.00; p=0.99; $\omega^2$=-0.022 | F(1,37)=2.36; p=0.13; $\omega^2$=0.006 | t(17)=0.74; p=0.47; d=0.17 | t(21)=-1.49; p=0.30; d=0.32 | t(37.79)=0.61; p=0.55; d=0.19 | t(37.70)=-0.65; p=0.55; d=0.21 |

*Table 3: Summary of analyses of variance and corresponding posthoc t-tests for perceived stress levels, heart rate, blood pressure, pain thresholds, pain tolerance and suprathreshold pain intensity and pain unpleasantness ratings. The table depicts the main effect of time (pre and post or pre and during the stressor respectively) and group (healthy controls (HC) and patients with chronic musculoskelettal pain (CMP)) and the time*group interaction effects. Posthoc tests were carried out for pre-post comparisons within each group and for HC-CMP comparisons before and after the stressor. Significant effects are highlighted in bold; NRS: Numeric Rating Scale*

## Discussion and conclusions

This study investigated if SIA is reduced in patients with CMP. We found increased pain thresholds after the stressor in controls, but not patients with CMP. Our secondary outcomes showed no change in pain tolerance or pain unpleasantness ratings for either group. However, pain intensity ratings increased after the stressor in patients with CMP, but not controls. An exploratory analysis of the spatial extent of the clinical pain showed no association with stress-related changes in primary or secondary outcomes.

*Distinct effects of stress on pain thresholds in pain patients and healthy individuals*

As expected, we found that the increase of pain thresholds after an acute stressor, i.e. SIA, is absent in patients with CMP compared to healthy controls. Patients with CMP perceive stress as an aggravating factor for their pain (Okifuji and Turk 2002) and the reduced capacity to downregulate pain, i.e. SIA, may explain why chronic pain persists in patients, whereas controls recover from acute pain episodes. This is also in line with studies showing reduced downregulation of experimental pain in patients with CMP in other experimental paradigms of pain inhibition such as conditioned pain modulation (O'Brien et al., 2018) or exercise induced hypoalgesia (Rice et al., 2019).

*Effects of stress on pain intensity ratings, pain unpleasantness ratings and pain tolerance in pain patients and healthy individuals*

Although prior studies using the same stressor found SIA but no SIH responses in healthy subjects (Flor et al., 2002; Yilmaz et al., 2010), we observed no such effect on suprathreshold pain intensity ratings, pain unpleasantness ratings, or pain tolerance. Previous studies investigated a younger control sample with an average age of about thirty (Flor et al., 2002; Flor and Grüsser 1999) to forty (Yilmaz et al., 2010) years. Two previous meta-analyses on conditioned pain modulation showed that pain-inhibition decreases with age, with absent pain-inhibition at an age above forty (Hackett et al., 2020; Lewis et al., 2012). With only 2 patients and 2 controls under the age of 40 we were not able to analyze this effect in detail, but we report that younger controls show stronger reduction of pain unpleasantness ratings after the stressor than older controls and that this effect is absent in patients with CMP (see Supplementary Figure 1).

Further, pain-inhibitory effects of different outcome parameters were previously reported as statistically independent (Nahman-Averbuch et al., 2013). This may be related to different peripheral nociceptive signaling pathways with differential sensitivity to thermal, mechanical or chemical stimulus modalities. In particular, heat and mechanical sensitivity of primate nociceptive afferents are inversely related (Treede et al., 1998; Treede et al., 1995), and mechanically sensitive nociceptors tend to have larger axons and lower electrical thresholds (Weidner et al., 1999). Thus, our findings are likely to be predictive for mechanical pain but less so for heat pain. This is in line with the finding of impaired conditioned pain modulation for electrical, but not heat pain thresholds (Lautenbacher and Rollman 1997). Further, suprathreshold pain ratings may be better predictors of chronicity than threshold measures (Granot et al., 2003) and showed large effect sizes (hedges g: 0.90) for deficient conditioned pain modulation in patients with chronic pain, whereas threshold measures yielded medium effect sizes (hedges g: 0.61) (Lewis et al., 2012). The only other study investigating the effect of the same stressor on suprathreshold pain ratings used mechanical test stimuli (Yilmaz et al., 2010) instead of electric test stimuli and showed decreased pain intensity and unpleasantness ratings after the stressor. Taken together, this indicates that measures of suprathreshold mechanical pain sensitivity may be more sensitive to stress-induced changes.

The lack of hypoalgesic effects on secondary outcomes may therefore be related to the age of our sample and the type of outcome measures. Nevertheless, we found indications of SIH, i.e. an increase in perceived pain intensity but not pain unpleasantness ratings after the stressor, in patients with CMP. Previously, perceived pain intensity, but not pain unpleasantness was modulated by the manipulation of attentional states, while the manipulation of emotional states primarily affected perceived pain unpleasantness (Villemure et al., 2003). Further, distinct brain pathways were found for the attentional effects on pain intensity and the effects of emotional modulation of pain unpleasantness (Villemure and Bushnell 2009). In patients with CMP, but not controls, stress may therefore have caused a narrowing of the attentional focus (Petrac et al., 2009) on painful stimulation (Broadbent et al., 2021) and thereby increased the perception of the sensory, but not affective component of pain.

Pain-related sensitization rather than habituation to repetitive painful stimulation has previously been reported in musculoskeletal pain patients (Diers et al., 2011) and hyperalgesia instead of analgesia has previously been observed in patients with fibromyalgia in paradigms of exercise-induced analgesia (Staud et al., 2005) and conditioned pain modulation (Potvin and Marchand 2016) where pain ratings in response to heat pain were assessed. Our data thus confirm and extend these prior findings to electrical stimulation and a cognitive stressor.

*Autonomic stress responsivity and its role in stress-induced pain modulation*

In healthy individuals, the responsiveness of the autonomic nervous system is an important component of the pain regulatory process and increased blood pressure levels are usually associated with greater pain inhibition (Hallman and Lyskov 2012). On the other hand, patients with chronic pain previously showed dysregulated stress systems such as a dysfunctional hypothalamic–pituitary–adrenal axis and autonomic nervous system dysregulation (Mostoufi et al., 2012; Walker et al., 2017; Woda et al., 2016). Further, persons with chronic low back pain previously displayed hyperalgesic effects of increased blood pressure, while the opposite relationship was reported for controls (Bruehl and Chung 2004). In the current study, patients with CMP showed stronger blood pressure increases during stress compared to controls. While findings on the autonomic nervous system in patients with CMP are contradictory, with decreased blood pressure reactivity after a cognitive stress task (Del Paso et al., 2010) or no altered blood pressure reactivity after a reaction time task (Nilsen et al., 2007), a similar stress task as in the current study also elicited elevated blood pressure responses in patients with fibromyalgia compared to controls (Thieme et al., 2015). We tested if this increased blood pressure reactivity also moderated the effects of stress on pain in the current study, but found no such effect. While the lack of significant increase in blood pressure in the control group could explain why we did not find this interaction effect, this group still showed stress-induced increases of pain thresholds. We therefore believe that our results rather indicate that the effects of this specific stressor on pain may not be mediated by the autonomic nervous system.

*Spatial pain extent and stress-induced pain modulation*

Our exploratory analysis did not confirm the hypothesis that patients with reduced SIA and enhanced pain ratings show a higher spatial pain extent rather than localized pain. Both SIA (Butler and Finn 2009) and SIH (Jennings et al., 2014) have been shown to be mediated by central mechanisms, which induce pain inhibition or facilitation via modulation of pathways descending from the periaqueductal grey and the rostral ventromedial medulla. It is therefore plausible that both, localized and widespread clinical pain, are affected by deficiencies in these mechanisms. However, we assessed a rather chronic sample, where initial differences between these groups may have subsided. Future studies should therefore investigate whether the decreased pain inhibition in patients with widespread pain that was previously found (Gerhardt et al., 2017; Woda et al., 2013), might be a function of pain chronicity, rather than extent alone.

*Limitations*

First, there is a high variability in the employed experimental paradigms in studies on stress-induced pain modulation (for an overview on experimental paradigms on SIA see (Butler and Finn 2009) and (Imbe et al., 2006; Jennings et al., 2014) for SIH). Especially in animal models of SIA, other types of stressors such as exposure to painful stimulation have been employed. Therefore, our results cannot be generalized to all SIA paradigms, as differences in experimental parameters may account for differences in the effects.

Second, our sample of patients with CMP included patients with localized and widespread pain and our groups were not perfectly matched for gender. Heterogeneous samples may increase the reliability of research findings (Lakes 2013) and we did not find a significant relationship between gender, spatial extent of the clinical pain and SIA or SIH. We therefore believe that our findings are independent of spatial pain extent and gender. However, there may be subgroups of patients who show only changes in either SIA or SIH and our sample size did not allow for further examination of such groups.

Third, we did not assess the effects of ongoing musculoskeletal pain on the effects of the stressor. This may be an important moderator of the differences between patients and controls, especially because endogenous pain inhibition was shown to differentially modulate ongoing and experimental pain (Witting et al., 2003). In addition.

ongoing pain demands attention (Eccleston and Crombez 1999) and different attentional states may modulate experimental and ongoing pain (Villemure and Bushnell 2002).

Fourth, the order of our outcome assessments was reversed after the stressor. This could pose a potential bias of the post-threshold measurements due to habituation and/or sensitization after applying the longer, suprathreshold pain testing blocks.

Fifth, we did not assess anticipated stress or expectations concerning the effects of the stress manipulation on pain outcomes. Previous work showed that expectations affect pain perception (Dannecker et al., 2003) as well as pain modulation (Benedetti et al., 1999; Price et al., 1999; Tracey 2010) and also that pain anticipation increases neural responses in the ACC (Koyama et al., 2005; Ploghaus et al., 2003; Sawamoto et al., 2000), which is involved in descending inhibition (Yilmaz et al., 2010), but also facilitation (Zhang et al., 2005) of pain by acute stress and shows exacerbated responses to acute stress after prior chronic stress exposure (Fee et al., 2020).

Sixth, in our patient cohort only two patients reported previous use of medication. This is low compared to other samples of patients with CMP and might indicate a sample of patients with mild to moderate musculoskeletal pain (consistent with the mean rating of 4.9/10).

*Conclusion*

The present study shows novel evidence of reduced SIA in patients with CMP for a purely cognitive experimental stressor. In daily life, such stressors are likely equally important as repeated muscle strains. Moreover, our electrical stimulation bypassed any peripheral sensitization of the transduction of natural stimuli (mechanical, heat), thus strengthening the interpretation that stress-induced modulations occur predominantly in the central nervous system. Further, patients with CMP displayed suprathreshold SIH, which was absent in controls. Currently, chronic pain and stress are often conceptualized as allostatic overload that causes bidirectional aggravation, underestimating the complexity of the relationship between different sources of stress and chronic pain (Abdallah and Geha 2017). Patients with CMP showed selective alterations in stress responsivity to personally relevant stress types (Flor et al., 1985) and our current findings indicate that particularly cognitive stress types may enhance pain perception in CMP. Future research should investigate if pain modulation by different stress types is a factor driving the development and maintenance of specific subgroups of CMP. Further, the role of additional stress response systems, such as the hypo-thalamic-pituitary-adrenal axis, and the neuronal and neurochemical mechanisms underlying these maladaptive effects of stress should be explored. A further open question is if our effects are specific for stress or related to attention, expectation, negative emotions and mood.

## Acknowledgments

This work was supported by grants of the Deutsche Forschungsgemeinschaft to FN and HF (SFB1158/B03), to FN (NE 1383/14-1), RDT (SFB1158/S01 and B09) and the Bundesministerium für Bildung und Forschung (01EC1940B) to HF and ML.

## Author contribution

ML, SK, PS, SSH, FN & HF designed the study; ML analyzed the data and wrote the paper; PS acquired the data, SSH, SK, KU, FN, RDT & HF revised the paper, FN, RDT & HF acquired the funding. All authors contributed to the interpretation of the data, revised the manuscript, and approved the final version of the manuscript.

## Data availability statement

The data generated during and/or analyzed during the current study are available from the corresponding author on reasonable request.

## Declarations of interest

none

## References

Abdallah CG and Geha P. Chronic pain and chronic stress: two sides of the same coin? Chronic Stress 2017;1: 2470547017704763.

Albers C and Lakens D. When power analyses based on pilot data are biased: Inaccurate effect size estimators and follow-up bias. Journal of experimental social psychology 2018;74: 187-195.

Benedetti F, Arduino C, Amanzio M. Somatotopic activation of opioid systems by target-directed expectations of analgesia. Journal of Neuroscience 1999;19: 3639-3648.

Benjamini Y and Hochberg Y. Approach to Multiple Testing. Journal of the Royal Statistical Society Series B (Methodological) 1995;57: 289-300.

Bjelland I, Dahl A, Haug T, Neckelmann D. The validity of the Hospital Anxiety and Depression Scale. An updated literature review. Journal of Psychosomatic Research 2002;52: 69.

Broadbent P, Liossi C, Schoth DE. Attentional bias to somatosensory stimuli in chronic pain patients: a systematic review and meta-analysis. Pain 2021;162: 332-352.

Bruehl S and Chung OY. Interactions between the cardiovascular and pain regulatory systems: an updated review of mechanisms and possible alterations in chronic pain. Neuroscience & Biobehavioral Reviews 2004;28: 395-414.

Butler R and Finn D. Stress-induced analgesia. Progress in Neurobiology 2009;88: 184.
Cathcart S, Petkov J, Pritchard D. Effects of induced stress on experimental pain sensitivity in chronic tension-type headache sufferers. European journal of neurology 2008;15: 552-558.
Cathcart S, Petkov J, Winefield A, Lushington K, Rolan P. Central mechanisms of stress-induced headache. Cephalalgia: an International Journal of Headache 2010;30: 285.
Cathcart S, Winefield AH, Lushington K, Rolan P. Effect of mental stress on cold pain in chronic tension-type headache sufferers. The Journal of Headache and Pain 2009;10: 367.
Choi J, Chung M, Lee Y. Modulation of pain sensation by stress-related testosterone and cortisol. Anaesthesia 2012;67: 1146.
Cohen J. A power primer. Psychological bulletin 1992;112: 155.
Costa A, Smeraldi A, Tassorelli C, Greco R, Nappi G. Effects of acute and chronic restraint stress on nitroglycerin-induced hyperalgesia in rats. Neuroscience Letters 2005;383: 7.
Crettaz B, Marziniak M, Willeke P, Young P, Hellhammer D, Stumpf A, Burgmer M. Stress-Induced Allodynia–Evidence of Increased Pain Sensitivity in Healthy Humans and Patients with Chronic Pain after Experimentally Induced Psychosocial Stress. PLoS ONE 2013;8.
Dannecker EA, Price DD, Robinson ME. An examination of the relationships among recalled, expected, and actual intensity and unpleasantness of delayed onset muscle pain. The Journal of Pain 2003;4: 74-81.
Del Paso GAR, Garrido S, Pulgar A, Martín-Vázquez M, Duschek S. Aberrances in autonomic cardiovascular regulation in fibromyalgia syndrome and their relevance for clinical pain reports. Psychosomatic Medicine 2010;72: 462-470.
Diers M, Schley M, Rance M, Yilmaz P, Lauer L, Rukwied R, Schmelz M, Flor H. Differential central pain processing following repetitive intramuscular proton/prostaglandin $E_2$ injections in female fibromyalgia patients and healthy controls. European Journal of Pain (London, England) 2011;15: 716.
Düker H and Lienert G. Konzentrations-Leistungs-Test (KLT) Hogrefe. Göttingen 1959.
Eccleston C and Crombez G. Pain demands attention: A cognitive–affective model of the interruptive function of pain. Psychological bulletin 1999;125: 356.
Fanselow M. Conditioned fear-induced opiate analgesia: a competing motivational state theory of stress analgesia. Annals of the New York Academy of Sciences 1986;467: 40.
Faul F, Erdfelder E, Lang A-G, Buchner A. G* Power 3: A flexible statistical power analysis program for the social, behavioral, and biomedical sciences. Behavior Research Methods 2007;39: 175-191.
Fee C, Prevot T, Misquitta K, Banasr M, Sibille E. Chronic stress-induced behaviors correlate with exacerbated acute stress-induced cingulate cortex and ventral hippocampus activation. Neuroscience 2020.
Ferreira DMAO, Costa YM, Bonjardim LR, Conti PCR. Effects of acute mental stress on conditioned pain modulation in temporomandibular disorders patients and healthy individuals. Journal of Applied Oral Science 2021;29.
Ferreira DMAO, Costa YM, de Quevedo HM, Bonjardim LR, Conti PCR. Experimental Psychological Stress on Quantitative Sensory Testing Response in Patients with Temporomandibular Disorders. Journal of oral & facial pain and headache 2018;32: 428–435.
Flor H, Behle D, Birbaumer N. Assessment of pain-related cognitions in chronic pain patients. Behaviour Research and Therapy 1993;31: 63.
Flor H, Birbaumer N, Schulz R, Grüsser S, Mucha R. Pavlovian conditioning of opioid and nonopioid pain inhibitory mechanisms in humans. European Journal of Pain (London, England) 2002;6: 395.
Flor H and Grüsser S. Conditioned stress-induced analgesia in humans. European Journal of Pain (London, England) 1999;3: 317.
Flor H, Rudy T, Birbaumer N, Streit B, Schugens M. The applicability of the West Haven-Yale multidimensional pain inventory in German-speaking countries. Data on the reliability and validity of the MPI-D. Schmerz 1990;4: 82.

Flor H, Turk DC, Birbaumer N. Assessment of stress-related psychophysiological reactions in chronic back pain patients. Journal of consulting and clinical psychology 1985;53: 354.

Fydrich T, Renneberg B, Schmitz B, Wittchen H.SKID II: Strukturiertes Klinisches Interview für DSM-IV (SKID-II), Achse II: Persönlichkeitsstörungen. [Structured Clinical Interview for DSM-IV (SCID), Axis 2: Personality Disorders] Eine deutschsprachige, erweiterte Bearbeitung der amerikanischen Originalversion des SCID-II von First MB, Spitzer RL, Gibbon M, Williams JBW. Benjamin L (Version 3/96). Göttingen, Hogrefe; 1997.

Gerhardt A, Eich W, Treede R, Tesarz J. Conditioned pain modulation in patients with nonspecific chronic back pain with chronic local pain, chronic widespread pain, and fibromyalgia. Pain 2017;158: 430.

Gerhardt A, Hartmann M, Blumenstiel K, Tesarz J, Eich W. The prevalence rate and the role of the spatial extent of pain in nonspecific chronic back pain—a population-based study in the south-west of Germany. Pain Medicine 2014;15: 1200-1210.

Granot M, Lowenstein L, Yarnitsky D, Tamir A, Zimmer EZ. Postcesarean section pain prediction by preoperative experimental pain assessment. Anesthesiology: The Journal of the American Society of Anesthesiologists 2003;98: 1422-1426.

Grosser T, Ricciotti E, FitzGerald GA. The cardiovascular pharmacology of nonsteroidal anti-inflammatory drugs. Trends in pharmacological sciences 2017;38: 733-748.

Hackett J, Naugle KE, Naugle KM. The decline of endogenous pain modulation with aging: a meta-analysis of temporal summation and conditioned pain modulation. The Journal of Pain 2020;21: 514-528.

Hallman D and Lyskov E. Autonomic Regulation in Musculoskeletal Pain. In: Pain in Perspective.London, United Kingdom: IntechOpen; 2012.

Imbe H, Iwai-Liao Y, Senba E. Stress-induced hyperalgesia: animal models and putative mechanisms. Frontiers in Bioscience: a Journal and Virtual Library 2006;11: 2179.

Inui K, Tran T, Hoshiyama M, Kakigi R. Preferential stimulation of Adelta fibers by intra-epidermal needle electrode in humans. Pain 2002;96: 247.

Jennings E, Okine B, Roche M, Finn D. Stress-induced hyperalgesia. Progress in Neurobiology 2014;121: 1.

Kerns R, Turk D, Rudy T. The West Haven-Yale Multidimensional Pain Inventory (WHYMPI). Pain 1985;23: 345.

Koyama T, McHaffie JG, Laurienti PJ, Coghill RC. The subjective experience of pain: where expectations become reality. Proceedings of the National Academy of Sciences 2005;102: 12950-12955.

Lacourt TE, Houtveen JH, van Doornen LJ. Experimental pressure-pain assessments: test–retest reliability, convergence and dimensionality. Scandinavian Journal of Pain 2012;3: 31-37.

Lakens D. Calculating and reporting effect sizes to facilitate cumulative science: a practical primer for t-tests and ANOVAs. Frontiers in psychology 2013;4: 863.

Lakes KD. Restricted sample variance reduces generalizability. Psychological assessment 2013;25: 643.

Lautenbacher S and Rollman G. Possible deficiencies of pain modulation in fibromyalgia. The Clinical journal of pain 1997;13: 189.

Lewis G, Rice D, McNair P. Conditioned pain modulation in populations with chronic pain: a systematic review and meta-analysis. The Journal of Pain 2012;13: 936-944.

Lewis J, Cannon J, Chudler E, Liebeskind J. Effects of naloxone and hypophysectomy on electroconvulsive shock-induced analgesia. Brain Research 1981;208: 230.

López-López A, Matías-Pompa B, Fernández-Carnero J, Gil-Martínez A, Alonso-Fernández M, Alonso Perez J, González Gutierrez J. Blunted Pain Modulation Response to Induced Stress in Women with Fibromyalgia with and without Posttraumatic Stress Disorder Comorbidity: New Evidence of Hypo-Reactivity to Stress in Fibromyalgia? Behavioral Medicine 2021;47: 311-323.

Lukesch H and Mayrhofer S. KLT-R. Revidierte Fassung des Konzentrations-Leistungs-Test von H. Düker & GA Lienert. 2001.
Mandelli M, Tognoni G, Garattini S. Clinical pharmacokinetics of diazepam. Clinical pharmacokinetics 1978;3: 72-91.
Mostoufi S, Afari N, Ahumada S, Reis V, Wetherell J. Health and distress predictors of heart rate variability in fibromyalgia and other forms of chronic pain. Journal of Psychosomatic Research 2012;72: 39.
Nagel B, Gerbershagen H, Lindena G, Pfingsten M. Entwicklung und empirische Überprüfung des Deutschen Schmerzfragebogens der DGSS. Der Schmerz 2002;16: 263-270.
Nahman-Averbuch H, Yarnitsky D, Granovsky Y, Gerber E, Dagul P, Granot M. The role of stimulation parameters on the conditioned pain modulation response. Scandinavian Journal of Pain 2013;4: 10-14.
Nilsen KB, Sand T, Westgaard RH, Stovner LJ, White LR, Leistad RB, Helde G, Rø M. Autonomic activation and pain in response to low-grade mental stress in fibromyalgia and shoulder/neck pain patients. European Journal of Pain 2007;11: 743-755.
O'Brien AT, Deitos A, Pego YT, Fregni F, Carrillo-de-la-Peña MT. Defective Endogenous Pain Modulation in Fibromyalgia: A Meta-Analysis of Temporal Summation and Conditioned Pain Modulation Paradigms. The Journal of Pain 2018.
Okifuji A and Turk D. Stress and psychophysiological dysregulation in patients with fibromyalgia syndrome. Applied Psychophysiology and Biofeedback 2002;27: 129.
Petrac D, Bedwell J, Renk K, Orem D, Sims V. Differential relationship of recent self-reported stress and acute anxiety with divided attention performance. Stress 2009;12: 313-319.
Ploghaus A, Becerra L, Borras C, Borsook D. Neural circuitry underlying pain modulation: expectation, hypnosis, placebo. Trends in cognitive sciences 2003;7: 197-200.
Potvin S and Marchand S. Pain facilitation and pain inhibition during conditioned pain modulation in fibromyalgia and in healthy controls. Pain 2016;157: 1704-1710.
Price DD, Milling LS, Kirsch I, Duff A, Montgomery GH, Nicholls SS. An analysis of factors that contribute to the magnitude of placebo analgesia in an experimental paradigm. Pain 1999;83: 147-156.
R_Core_Team.R: A language and environment for statistical computing. Vienna, Austria: R Foundation for Statistical Computing; 2017.
Reinhardt T, Kleindienst N, Treede R, Bohus M, Schmahl C. Individual modulation of pain sensitivity under stress. Pain Medicine 2013;14: 676.
Rice D, Nijs J, Kosek E, Wideman T, Hasenbring MI, Koltyn K, Graven-Nielsen T, Polli A. Exercise-induced hypoalgesia in pain-free and chronic pain populations: state of the art and future directions. The Journal of Pain 2019;20: 1249-1266.
Rosenstiel A and Keefe F. The use of coping strategies in chronic low back pain patients: relationship to patient characteristics and current adjustment. Pain 1983;17: 33.
Sawamoto N, Honda M, Okada T, Hanakawa T, Kanda M, Fukuyama H, Konishi J, Shibasaki H. Expectation of pain enhances responses to nonpainful somatosensory stimulation in the anterior cingulate cortex and parietal operculum/posterior insula: an event-related functional magnetic resonance imaging study. Journal of Neuroscience 2000;20: 7438-7445.
Schulz P, Schlotz W, Becker P. Trierer Inventar zum chronischen Stress: TICS Hogrefe. 2004.
Smith B, Penny K, Purves A, Munro C, Wilson B, Grimshaw J, Chambers W, Smith W. The Chronic Pain Grade questionnaire: validation and reliability in postal research. Pain 1997;71: 141.
Snaith R and Zigmond A. The Hospital Anxiey and Depression Scale with the Irritability-depression-anxiety Scale and the Leeds Situational Anxiety Scale: Manual Nfer-Nelson. 1994.
Staud R, Robinson M, Price D. Isometric exercise has opposite effects on central pain mechanisms in fibromyalgia patients compared to normal controls. Pain 2005;118: 176.
Thieme K, Turk D, Gracely R, Maixner W, Flor H. The relationship among psychological and psychophysiological characteristics of fibromyalgia patients. The Journal of Pain 2015;16: 186.

Tracey I. Getting the pain you expect: mechanisms of placebo, nocebo and reappraisal effects in humans. Nature medicine 2010;16: 1277-1283.
Treede R, Meyer R, Campbell J. Myelinated mechanically insensitive afferents from monkey hairy skin: heat-response properties. Journal of neurophysiology 1998;80: 1082-1093.
Treede R, Meyer R, Raja S, Campbell J. Evidence for two different heat transduction mechanisms in nociceptive primary afferents innervating monkey skin. The Journal of physiology 1995;483: 747-758.
Vanegas H, Vazquez E, Tortorici V. NSAIDs, opioids, cannabinoids and the control of pain by the central nervous system. Pharmaceuticals 2010;3: 1335-1347.
Villemure C and Bushnell CM. Cognitive modulation of pain: how do attention and emotion influence pain processing? Pain 2002;95: 195-199.
Villemure C and Bushnell MC. Mood influences supraspinal pain processing separately from attention. Journal of Neuroscience 2009;29: 705-715.
Villemure C, Slotnick BM, Bushnell MC. Effects of odors on pain perception: deciphering the roles of emotion and attention. Pain 2003;106: 101-108.
Von Korff M, Ormel J, Keefe F, Dworkin S. Grading the severity of chronic pain. Pain 1992;50: 133.
Walker LS, Stone AL, Smith CA, Bruehl S, Garber J, Puzanovova M, Diedrich A. Interacting influences of gender and chronic pain status on parasympathetically-mediated heart rate variability in adolescents and young adults. Pain 2017;158: 1509.
Weidner C, Schmelz M, Schmidt R, Hansson B, Handwerker H, Torebjörk H. Functional attributes discriminating mechano-insensitive and mechano-responsive C nociceptors in human skin. Journal of Neuroscience 1999;19: 10184-10190.
Willer JC, Dehen H, Cambier J. Stress-induced analgesia in humans: endogenous opioids and naloxone-reversible depression of pain reflexes. Science 1981;212: 689-691.
Wittchen H, Zaudig M, Fydrich T. Strukturiertes Klinisches Interview für DSM-IV (SKID), Achse 1 [Structured Clinical Interview for DSM-IV (SCID), Axis 1 Disorders]. Hogrefe, Göttingen 1997.
Witting N, Svensson P, Jensen TS. Differential recruitment of endogenous pain inhibitory systems in neuropathic pain patients. PAIN® 2003;103: 75-81.
Woda A, L'heveder G, Ouchchane L, Bodéré C. Effect of experimental stress in 2 different pain conditions affecting the facial muscles. The Journal of Pain 2013;14: 455.
Woda A, Picard P, Dutheil F. Dysfunctional stress responses in chronic pain. Psychoneuroendocrinology 2016;71: 127.
Wolfe F, Clauw D, Fitzcharles M, Goldenberg D, Katz R, Mease P, Russell A, Russell I, Winfield J, Yunus M. The American College of Rheumatology preliminary diagnostic criteria for fibromyalgia and measurement of symptom severity. Arthritis Care & Research 2010;62: 600.
Wolfe F, Clauw DJ, Fitzcharles M-A, Goldenberg DL, Häuser W, Katz RS, Mease P, Russell AS, Russell IJ, Winfield JB. Fibromyalgia criteria and severity scales for clinical and epidemiological studies: a modification of the ACR Preliminary Diagnostic Criteria for Fibromyalgia. The Journal of Rheumatology 2011;38: 1113-1122.
Yilmaz P, Diers M, Diener S, Rance M, Wessa M, Flor H. Brain correlates of stress-induced analgesia. Pain 2010;151: 522.
Zhang L, Zhang Y, Zhao Z. Anterior cingulate cortex contributes to the descending facilitatory modulation of pain via dorsal reticular nucleus. The European journal of neuroscience 2005;22: 1141.

## Supplement

| | Time | group | Time*group | Time*gender | Time*age | Time*gender*group | Time*age*group |
|---|---|---|---|---|---|---|---|
| **Pain Threshold (in mA)** | **F(1,37)=4.55; p=0.04; $\omega^2$=0.017** | **F(1,37)=6.69; p=0.01; $\omega^2$=0.103** | **F(1,37)=6.48; p=0.02; $\omega^2$=0.027** | F(1,37)=2.30; p=0.14; $\omega^2$=0.006 | F(1,37)=1.32; p=0.26; $\omega^2$=0.002 | F(1,37)=0.30; p=0.58; $\omega^2$=-0.003 | F(1,37)=0.27; p=0.61; $\omega^2$=-0.004 |
| **Pain Tolerance (in mA)** | F(1,37)=0.06; p=0.81; $\omega^2$=-0.006 | **F(1,37)=7.24; p=0.01; $\omega^2$=0.111** | F(1,37)=0.12; p=0.73; $\omega^2$=-0.006 | F(1,37)=0.003; p=0.96; $\omega^2$=-0.006 | F(1,37)=1.31; p=0.26; $\omega^2$=-0.002 | F(1,37)=0.45; p=0.51; $\omega^2$=-0.003 | F(1,37)=0.67; p=0.42; $\omega^2$=-0.002 |
| **Perceived pain intensity (NRS)** | **F(1,37)=4.80; p=0.04; $\omega^2$=0.030** | F(1,37)=2.78; p=0.11; $\omega^2$=0.028 | **F(1,37)=7.40; p=0.01; $\omega^2$=0.050** | F(1,37)=3.51; p=0.07; $\omega^2$=0.020 | F(1,37)=0.002; p=0.96; $\omega^2$=-0.008 | F(1,37)=0.006; p=0.94; $\omega^2$=-0.008 | F(1,37)=0.61; p=0.44; $\omega^2$=-0.003 |
| **Perceived pain unpleasantness (NRS)** | F(1,37)=0.15; p=0.70; $\omega^2$=-0.003 | F(1,37)=0.000; p=0.99; $\omega^2$=-0.022 | F(1,37)=2.42; p=0.13; $\omega^2$=0.004 | F(1,37)=1.91; p=0.18; $\omega^2$=-0.001 | F(1,37)=0.67; p=0.42; $\omega^2$=0.000 | F(1,37)=1.03; p=0.32; $\omega^2$=0.000 | **F(1,37)=4.31; p=0.047; $\omega^2$=0.010** |

*Supplementary Table 1: Summary of analyses of variance for pain thresholds, pain tolerance and suprathreshold pain intensity and pain unpleasantness ratings, including gender and age as covariates. The table depicts the main effect of time (pre and post or pre and during the stressor respectively), group (healthy controls (HC) and patients with chronic pain (CMP)), their interaction effects and all interactions of the covariates with these effects. Significant effects are highlighted in bold; NRS: Numeric Rating Scale, mA: milli Ampere.*

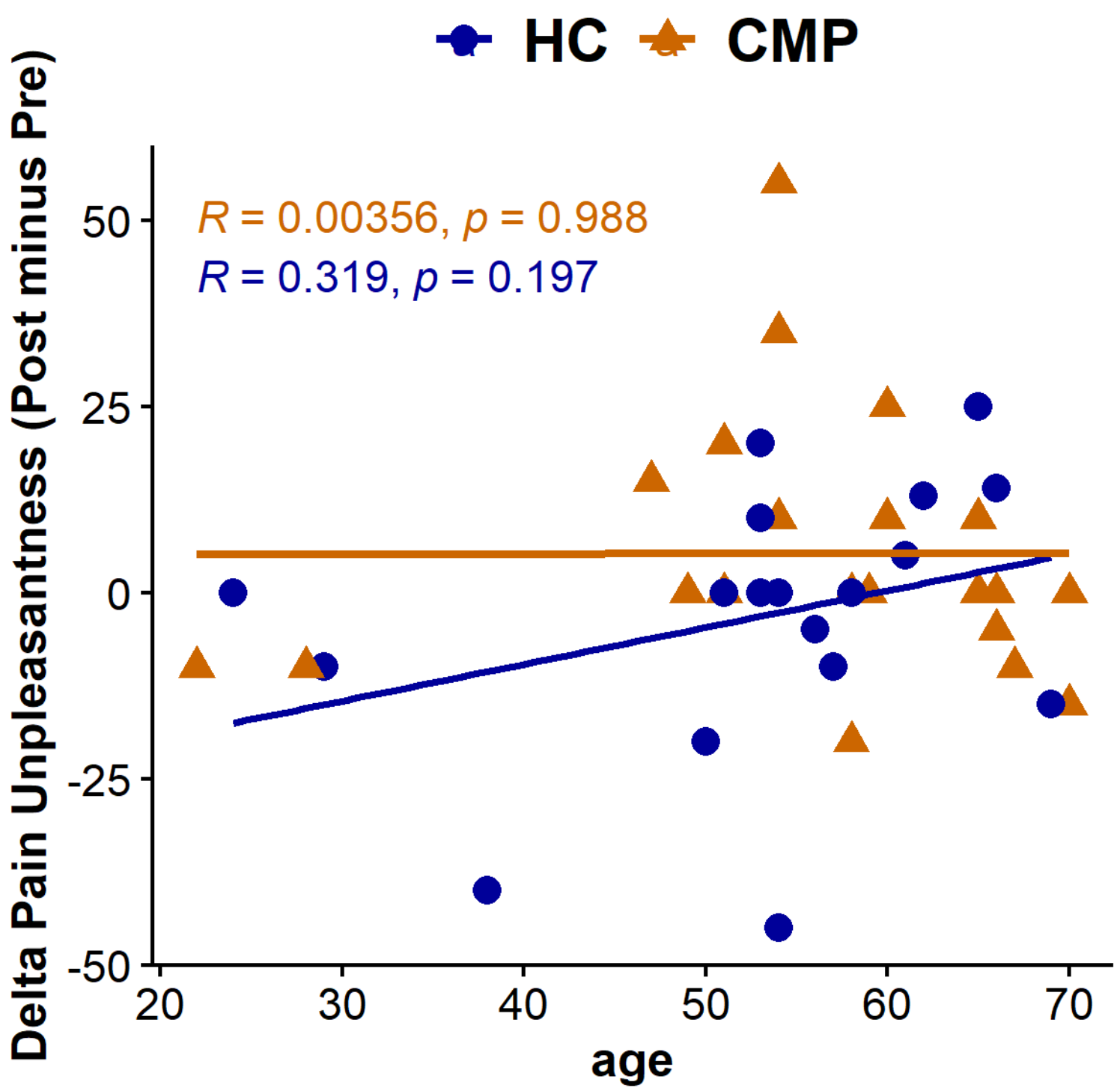

*Supplementary Figure 1: Depiction of the significant interaction effect of the factors time, group and age on pain unpleasantness ratings. Single subjects are depicted as orange triangles (patients with chronic musculoskeletal pain, CMP) and blue circles (healthy controls, HC). Regression Lines and Pearson's correlation coefficients between age and pain unpleasantness ratings are depicted for each group (blue and orange) and timepoint (pre and post) separately. The graph shows that in HC pain unpleasantness ratings after the stressor decrease more in younger subjects, while no such effect is shown for patients with CMP. NRS: Numeric Rating Scale.*

| | Time | group | Time* group | Time* gender | Time* age | Time* BP systolic | Time* gender* group | Time* age* group | Time* group* BP systolic |
|---|---|---|---|---|---|---|---|---|---|
| **Pain Threshold (in mA)** | F(1,37)=3.83; p=0.066; $\omega^2$=0.016 | **F(1,37)=5.39; p=0.026; $\omega^2$=0.100** | **F(1,37)=4.49; p=0.048; $\omega^2$=0.020** | F(1,37)=2.03; p=0.171; $\omega^2$=0.006 | F(1,37)=1.11; p=0.307; $\omega^2$=0.001 | F(1,37)=1.64; p=0.216; $\omega^2$=0.004 | F(1,37)=0.06; p=0.807; $\omega^2$=-0.005 | F(1,37)=0.25; p=0.620; $\omega^2$=-0.004 | F(1,37)=0.50; p=0.488; $\omega^2$=-0.003 |
| **Pain Tolerance (in mA)** | F(1,37)=0.07; p=0.800; $\omega^2$=-0.006 | **F(1,37)=13.04; p=0.002; $\omega^2$=0.120** | F(1,37)=0.13; p=0.718; $\omega^2$=-0.005 | F(1,37)=0.16; p=0.697; $\omega^2$=-0.005 | F(1,37)=3.02; p=0.099; $\omega^2$=0.012 | F(1,37)=1.33; p=0.264; $\omega^2$=0.002 | F(1,37)=0.11; p=0.745; $\omega^2$=-0.005 | F(1,37)=0.12; p=0.735; $\omega^2$=-0.005 | F(1,37)=1.40; p=0.251; $\omega^2$=0.002 |
| **Perceived pain intensity (NRS)** | **F(1,37)=4.75; p=0.043; $\omega^2$=0.030** | F(1,37)=2.50; p=0.130; $\omega^2$=0.027 | **F(1,37)=8.11; p=0.011; $\omega^2$=0.056** | F(1,37)=4.09; p=0.058; $\omega^2$=0.024 | F(1,37)=0.08; p=0.780; $\omega^2$=-0.007 | F(1,37)=3.33; p=0.085; $\omega^2$=0.018 | F(1,37)=1.10; p=0.308; $\omega^2$=0.001 | F(1,37)=0.61; p=0.445; $\omega^2$=-0.003 | F(1,37)=0.00; p=0.992; $\omega^2$=-0.008 |
| **Perceived pain unpleasantness (NRS)** | F(1,37)=0.15; p=0.705; $\omega^2$= -0.003 | F(1,37)=0.00; p=0.99; $\omega^2$=-0.026 | F(1,37)=3.93; p=0.063; $\omega^2$=0.009 | F(1,37)=2.74; p=0.115; $\omega^2$=0.005 | F(1,37)=1.52; p=0.234; $\omega^2$=0.002 | F(1,37)=0.13; p=0.723; $\omega^2$=-0.003 | F(1,37)=0.42; p=0.526; $\omega^2$=-0.002 | F(1,37)=3.70; p=0.070; $\omega^2$=0.008 | F(1,37)=1.62; p=0.220; $\omega^2$=0.002 |

*Supplementary Table 2: Summary of analyses of variance for pain thresholds, pain tolerance and suprathreshold pain intensity and pain unpleasantness ratings, including gender and age as covariates. The table depicts the main effect of time (pre and post or pre and during the stressor respectively), group (healthy controls (HC) and patients with chronic pain (CMP)), their interaction effects and all interactions of the covariates with these effects. Significant effects are highlighted in bold; NRS: Numeric Rating Scale, mA: milli Ampere.*

# S1

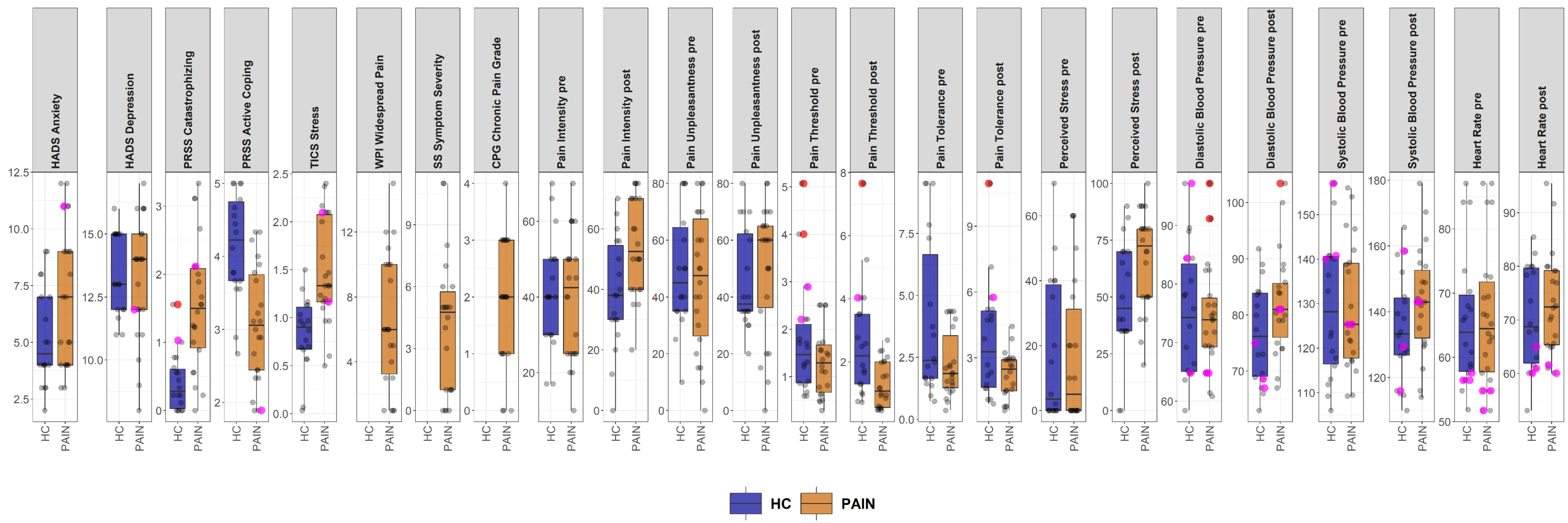

*Supplementary Figure 2: Distribution of original and imputed data is shown as boxplots and single data points. Thick horizontal bars represent the median, the histogram represents the first quartile above or below the median, the whiskers represent the second quartile above the median pain score. Red circles denote the presence of outlier(s) and magenta circles show imputed data. HC: Healthy Control; PAIN: Patients with chronic pain;*